\documentclass{aastex}
\usepackage{emulateapj5}
\usepackage{epsfig}
\usepackage{graphicx}
\begin{document}
\title{The detection of ultra-faint low surface brightness dwarf galaxies in the Virgo Cluster: a Probe of Dark Matter and Baryonic Physics}
\author{E. Giallongo$^1$, N. Menci$^1$, A. Grazian$^1$, R. Fassbender$^{1,2}$, A. Fontana$^1$, D. Paris$^1$, L. Pentericci$^1$}
\affil{$^1$INAF - Osservatorio Astronomico di Roma, via di Frascati
33, I-00040 Monteporzio, Italy\\
$^2$ Max-Planck-Institut f\"ur extraterrestrische Physik (MPE), Postfach 1312, Giessenbachstr.,  85741 Garching, Germany   \label{MPE}}

\smallskip

\begin{abstract}

We have discovered 11 ultra-faint  ($r\lesssim 22.1$) low surface brightness (LSB, central surface brightness $23\lesssim \mu_r\lesssim 26$) dwarf galaxy candidates in one deep Virgo field of just $576$ arcmin$^2$ obtained by the Large Binocular Camera (LBC) at the Large Binocular Telescope (LBT). Their association with the Virgo cluster is supported by their distinct position in the central surface brightness - total magnitude plane with respect to the background galaxies of similar total magnitude. They have typical absolute magnitudes and scale sizes, if at the distance of Virgo, in the range $-13\lesssim M_r\lesssim -9$ and $250\lesssim r_s\lesssim 850$ pc, respectively. Their colors are consistent with a gradually declining star formation history with a specific star formation rate of the order of $10^{-11}$ yr$^{-1}$, i.e. 10 times lower than that of main sequence star forming galaxies. They are older than the cluster formation age and appear regular in morphology. They represent the faintest extremes of the population of low luminosity LSB dwarfs that has been recently detected in wider surveys of the Virgo cluster. Thanks to the depth of our observations we are able to extend the Virgo luminosity function down to $M_r\sim -9.3$ (corresponding to total masses $M\sim 10^7$ M$_{\odot}$), finding an average faint-end slope $\alpha\simeq -1.4$. This relatively steep slope puts interesting constraints on the nature of the Dark Matter and in particular on warm Dark Matter (WDM) often invoked to solve the overprediction of the dwarf number density by the standard CDM scenario. We derive a lower limit on the WDM particle mass $>1.5$ keV.
\end{abstract}

\keywords{cosmology: observations --- galaxies: clusters: individual (Virgo)}

\section{Introduction}

The predicted over-abundance of galaxy satellites is a well known challenge to the standard Cold Dark Matter (CDM) scenario, and it appears to be related to the large small-scale power of the CDM power spectrum of density perturbations (see e.g., \citet{Klypin99}).
The inclusion of baryonic processes in the models can reduce this discrepancy (e.g., by suppressing  star formation through supernovae feedback or photoionizing UV background, \citet{Governato07}), although it is still difficult to maintain a simultaneous matching with other properties of satellite galaxies, like colors and star-formation histories. An interesting alternative is to assume that Dark Matter is composed of particles with mass $m_X\sim 1$ keV (warm Dark Matter, e.g., \citet{Sommer01}). The larger corresponding free-streaming length results in a strong suppression of the power spectrum at the dwarf galaxy scales (\citet{Bode01}). This effect reduces the expected abundance of dwarfs with respect to the predictions of CDM models, providing a better agreement with observations (e.g., \citet{Papastergis14}).

In this context, an attractive way to probe different DM cosmological scenarios is to look for  faint dwarf galaxies in the local universe and in particular in the Virgo cluster, which represents the densest nearby region. This large galaxy concentration allows us to probe the abundance of dwarf galaxies down to DM masses $\sim 10^7$ M$_{\odot}$ (e.g., \citet{Sabatini05}) where predictions of WDM and CDM diverge (\citet{Schneider12}). Nevertheless, the search for very faint dwarfs in Virgo is not straightforward, as it implies pushing the detection toward very low limits in central surface brightness, as first noted by \citet{Sandage85}. 

\citet{Impey88} found a population of blue LSB dwarfs in the Virgo cluster with $M_B<-11$ not detected in previous surveys. They stressed the effects of this new LSB dwarf population on the faint-end of the Virgo galaxy luminosity function (LF hereafter).

More recently \citet{Trentham02} performed a survey in five different local fields of varying galaxy density including the center of the Virgo cluster. They obtained relatively deep Subaru images in a small area of the Virgo cluster surrounding the M86 bright galaxy. They made also use of part of the 'VCC' sample of \cite{Binggeli85} to derive the bright end of the LF. Their new Subaru dwarf sample allowed the evaluation of the Virgo LF down to $M_r\sim -12$. Since the slope found at the faint end ($-1.2$) was significantly flatter than the CDM mass function slope ($-1.9$), they suggested a high degree of suppression of star formation in small DM halos. The suppressed luminosity of low-mass DM clumps would then explain the 
flat slope of the luminosity function.

\citet{Sabatini03} performed a wide survey (14 deg$^2$) down to $M_B\sim -10$ with the Isaac Newton Telscope (2.5 m). Despite some success in discovering faint LSB dwarfs, they could not reach a high level of completeness and their derived LF appears flat in the magnitude interval $-13<M_B<-11$. However, when combining their sample with the brighter Virgo cluster sample, they suggest a somewhat steeper faint end LF than previously found.

A recent investigation of the dwarf population using CFHT imaging has been performed by \citet{Lieder12}, reaching a completeness
limit of $M_V\sim -13$ in 3.75 deg$^2$. They found an average slope of the LF as steep as $\alpha=-1.5$ in the range $-19\lesssim M_V\lesssim -13$.

Very recently, \citet{Davies15} have provided a first analysis of the LSB dwarf population in a field of about 100 deg$^2$ from the Next Generation Virgo Survey (\citet{Ferrarese12}). This survey is larger with respect to the one by \citet{Sabatini03} although not deeper, reaching completeness at $g\sim 19$ (which approximately corresponds to $r\sim 18.5$ for an average $g-r\sim 0.5$). Their LF slope is consistent with previous values $\alpha \sim -1.35$.

To summarize, a population of faint LSB dwarfs in the Virgo cluster has emerged  from  imaging analysis, although the images were not deep enough to reach firm conclusions about the shape of the very faint-end of the Virgo LF.

In this paper we show the results of a {\it pilot} LSB dwarf search in a deep Virgo field. Although smaller than previous surveys, our observations are significantly deeper  thanks to the use of the Large Binocular Camera (LBC, \citet{Giallongo08}) of the Large Binocular Telescope.

The images were acquired by \citet{Lerchster11} (see also \citet{Fassbender11}) in the context of the study of the high redshift ($z\sim 1$) galaxy cluster XMMU J1230.3+1339, that is located in the projected region of the Virgo cluster including NGC 4477. Although the field of view covers an area of only 0.16 deg$^2$, it is particularly deep and provides a first estimate of the dwarf surface density at the faint-end of the galaxy LF in Virgo.

We adopt $\Omega_{\Lambda}=0.7$, $\Omega_{0}=0.3$, and $h=0.7$ in units of 100 km/s/Mpc. AB magnitudes have been adopted except for Figure 5.

\section{The LSB dwarf galaxy sample}

The images were obtained in the  U,B,r,i,z bands with exposure times ranging from 3.2 ks to 9 ks, and calibrated using  zeropoints derived from Landoldt standards or Sloan DR7 catalog \citet{Abazajian09}. The size of the field of view at full depth is $\sim 576$ arcmin$^2$. Further observational details can be found in \citet{Lerchster11}.

To detect faint LSB dwarfs and to confirm their association with the Virgo cluster we have used their distribution in the  $\mu_0-r$ plane, as previously done  by several authors (e.g. \citet{Conselice02}, \citet{Rines08} and reference therein). 
 We have performed the detection  in the $r$ band using SExtractor (\citet{Bertin96}) setting a surface brightness limit of $\mu \simeq 27.2$ at the 2$\sigma$ level and a minimum detection circular area of diameter 2.5 arcsec. SExtractor has also been used to derive a first estimate of the central surface brightness. We plot the galaxies found down to $r=23$ as a function of the central surface brightness in Figure 1. 
There are 11 galaxies that are well separated from the bulk of the galaxies in the catalog  (note that magnitudes and central surface brightnesses of the blue circles are more accurately derived from profile fitting, see below). A visual inspection has been executed to confirm the reality of such detections. All 11 LSB dwarf candidates are clearly extended sources with low surface brightness, and they are all detected in the full U,B,r,i,z multicolor dataset used. A further visual inspection over the whole image failed to detect additional candidates.

To perform a more quantitative selection, we have plotted in the same Figure 1 a straight line representing the LSB selection threshold used by \citet{Rines08} to separate Virgo galaxies from the background sample, extrapolated to our fainter magnitudes. 9 out of our 11 candidates are above or very close to their selection threshold, and the remaining 2 are in any case well separated by the locus of the background sample. We note that the \citet{Rines08} threshold is a robust but not complete selection criterion, since there is a small overlap between the two populations, as shown by the spectroscopic detection of a few Virgo galaxies below the selection threshold (i.e. at brighter central surface brightness with respect to the adopted line for a given total magnitude). 
As an independent check, we have also verified that dwarfs with similar characteristics are absolutely absent in our database of LBC fields even deeper than this (i.e., with exposure times $\sim 3-10$ hr and limiting magnitudes $r\sim 26-27$) obtained in different sky regions for a total area of $\sim 0.6$ deg$^2$ (\citet{Boutsia14}, Grazian et al. in preparation), lending further support to the association of the selected LSB dwarf candidates to the Virgo cluster.
For these reasons, in the following we will consider all the 11 LSB dwarfs as bona fide Virgo members.

The positions of all candidates in our image are shown in Figure 2 and their $r$-band morphologies are shown in Figure 3. In the latter, their low central surface brightness with respect to the background sources is clearly evident. In some cases compact sources are overlapping with our LSB dwarf candidates, possibly altering the estimate of their total magnitude and morphological parameters. 
To obtain cleaner measurements we have used the Galfit package  (\citet{Peng10}), a versatile software that allows radial profile modeling with different functions. We used S{\'e}rsic power-law profiles (\citet{Sersic68}), without truncation, usually adopted to fit spiral as well as spheroidal morphologies. In the fitting procedure we leave the S{\'e}rsic power-law index $n$, the scale radius $r_s$, the axial ratio, the total magnitude (or the central surface brightness) as free parameters. The De Vaucouleurs and exponential disk profiles are included as cases with S{\'e}rsic index $n=4$ or 1, respectively. The PSF profile was convolved with the intrinsic profiles for the fitting. Small, compact sources superimposed on our LSB dwarfs were simultaneously fit with the main LSB galaxy to remove their contribution. The most striking example is given by ``K'' galaxy which is contaminated by several compact  sources, and which is shown in more detail in Figure 4. In this specific case differences with respect to the Sextractor parameters can reach values of the order of 0.4 in total magnitude. In general, however, differences between Galfit and Sextractor total magnitudes or central surface brightnesses are no more than 0.15 mag or mag arcsec$^{-2}$, respectively.

The resulting parameters, including magnitudes and central surface brightnesses, for the LSB dwarf candidates are shown in Table 1. For most sources the central surface brightness in the $r$ band lies within the interval $\mu_0\sim 24-26$ and the S{\'e}rsic scale radius in the interval $r_s=3-10.7$ arcsec or equivalently $r_s=250-850$ pc adopting an angular scale of 80 pc arcsec$^{-1}$, that corresponds to a distance of the Virgo cluster of $16.5$ Mpc. Absolute magnitudes derived from the fit are in general fainter than $M_r>-13$ and reach values as faint as $M_r\sim -9.3$ in the $r$ band. Also, the S{\'e}rsic index $n$ is always $n\lesssim 1$ typical of high surface brightness spirals and dwarfs.  In general the axial ratio is larger than 0.6, and for $\sim 50$\% of the sample it is $\gtrsim 0.8$.  Formal statistical errors provided by Galfit are typically a few percent.

To estimate galaxy colors in a similar way, without the contribution from overlapping compact sources, we have repeated the same procedure in our U,B,i,z images. We have again used Galfit, fixing the radial profile obtained in the $r$ band image and leaving only the total magnitude free to change. In the estimate we assumed negligible color gradients along the galaxy profile. We have checked that the $\chi^2$ did not increase because of this assumption. The procedure is often needed in the  low signal-to-noise blue and UV images where only the central part of the galaxy profile 
emerges from the background noise. Results are shown in Table 1, where all the galaxies appear red with an average $\langle B-r\rangle \sim 0.8$.
To compare with the Sloan colors we fit the observed colors to theoretical spectral energy distributions (SED) using the procedure described below. We computed the Sloan $g$ magnitude from the resulting best--fitting spectrum, yielding an average $\langle g-r\rangle \sim 0.5$. 
The latter color is somewhat bluer than that of the brightest Virgo galaxies in our field ($g-r\sim 0.9, 0.8$ for NGC 4477 and 4479, respectively). The  LSB colors are also  bluer than predictions from CDM models ($g-r\sim 0.7$, \citet{Weinmann11}), which include specific prescriptions for star formation and feedback for simulated brighter Virgo dwarfs with $M_r<-15.5$. As suggested by the same authors, the model predicts redder colors mainly because it could overestimate the star formation quenching in Virgo dwarfs due to environmental effects (e.g., ram pressure stripping), required to make their number density consistent with the observations.

The LSB dwarfs found in our field are similar to dwarfs found, e.g., by \cite{Sabatini03}, \citet{Lieder12} or \citet{Davies15} but are considerably fainter.
In particular we have only one dwarf in common with \citet{Davies15}: in our data the "K" dwarf (LSBVCC025) appears more diffuse with a scale radius
twice larger. This is a combined effect of the larger depth of our data and of our Galfit analysis, which shows that appropriate photometry can be derived only after removing the faint background sources present inside the LSB profile (see Figure 4). For this reason, a direct comparison between the magnitudes derived by us and by \citet{Davies15} must take into account the contribution by the faint interlopers, which amounts to $\Delta r\sim -0.4$. By assuming an average $g-r\sim 0.4$ (as derived from our table 1 with appropriate interpolation and filter conversion) we obtain the value $g\sim 19.1$ as in \citet{Davies15}.
 
We have derived physical quantities for the 11 LSB dwarfs fitting the UBriz multicolor data to the synthetic SEDs based on a grid of models from the \citet{Bruzual03} (BC) spectral synthesis code. The synthetic spectra are characterized by exponentially declining star formation histories of timescale $\tau$ and by a set of ages, metallicities and dust extinctions (see, e.g., \citet{Grazian06}, \citet{Fontana06} for details). The redshift has been fixed to that of the Virgo cluster (peculiar velocities are irrelevant in this context).

The output of this SED fitting procedure is shown in  Table 1. All the galaxies are consistent with a stellar metallicity 0.2 solar and no dust. The resulting star formation rates (SFR) and stellar masses are indicative of old stellar populations with low SFR. In particular the ratio $\langle age/\tau\rangle \sim 3.6$ is indicative of a passively declining stellar population, although this ratio is not as large as in brighter ellipticals ($\langle age/\tau\rangle > 6$  \citet{Fontana09}).

Moreover, extrapolating  the linear relation between star formation and stellar mass found for the main sequence galaxies in the local universe (see  \citet{Elbaz11}) toward low masses, our dwarfs show average SFRs a factor 10 lower for a given stellar mass with respect to the linear relation, consistent with a quenched star formation phase with typical specific SFR $\sim 10^{-11}$ yr$^{-1}$.

\section{The Virgo luminosity function}  
The main outcome of the present study is the large number density of LSB dwarfs found in a single LBC pointing,  corresponding to  11 LSB Virgo candidates (after removing bright extended objects) in an effective field of 0.12 deg$^2$.  The derived projected number density can be compared with corresponding projected densities of brighter Virgo galaxies.  We use here the \citet{Trentham02} sample, whose projected LF is computed per square megaparsec, per magnitude and is normalized at a distance from the Virgo center (i.e. M87 position) of 200 kpc.  We split our small sample into two bins  (each 2 mags wide, one centered at $M_r=-12.5$ and one at at $M_r=-10.5$) that contain 4 and 7 candidates, respectively.

Adopting for Virgo the angular scale  of 80 pc arcsec$^{-1}$ we obtain a density of 200 and 350 objects Mpc$^{-2}$ mag$^{-1}$ in our two bins at the position of our field, which is about 1.3 deg or 340 kpc from M87. To rescale to 200 kpc for comparison with the \citet{Trentham02} LF we have adopted their projected radial density distribution of brighter Virgo galaxies. We derive an increase by a factor $\simeq 1.6$ from 340 to 200 kpc. The resulting projected densities of LSB dwarfs at 200 kpc scale to 320 and 560 Mpc$^{-2}$ mag$^{-1}$, respectively.

These two average densities derived in our relatively small field are shown in Figure 5 together with LF data adapted from \citet{Trentham02}. Under the hypothesis that the small LSB dwarfs are distributed following the smooth gravitational potential well of the Virgo cluster we assume that the derived density is a fair representation of the average LSB dwarf density at $\sim 300$ kpc from the Virgo center. We explicitly neglect any possible dwarf clustering or anti-clustering around the bright Virgo galaxies.
Our sample shows an excess at $M_r>-17$ compared to the LF of \citet{Trentham02} but it is difficult to ascertain whether this is due to a real deviation of the LF from a Schechter shape (\citet{Trentham02}) or to observational biases which produce deviations from a featureless faint-end power law (see, e.g., \citet{Lieder12}). In practice the LF is derived by connecting points from \citet{Trentham02} brighter than $M_r=-17$, which are mainly based on the VCC survey (\citet{Binggeli85}), with our two faint bins. The resulting slope is $\alpha\sim -1.4$ and it is shown in Figure 5. We also show in the same figure a flatter $-1.3$ and steeper $-1.5$ slope for comparison. Our value is consistent with very recent estimates derived from the overall Virgo region by \citet{Davies15} $\alpha\simeq -1.35$, in the $g$ band.  However, their slope evaluation is limited to $g\sim 18$ which corresponds to $r\sim 17.5$ assuming $g-r\sim 0.5$. Although statistically limited by the small LBC area, our complementary data allows us to probe the extension of the LF slope about 4 magnitudes fainter.

\section{The implications for Dark Matter scenarios}  

The presence of a significant density of faint dwarfs in a rich cluster like Virgo can put interesting constraints on the nature of Dark Matter and on the physical mechanisms regulating galaxy star formation activity. In DM scenarios, galaxies form at the center of DM halos collapsed in over-dense regions. Following the progressive inclusion of previously collapsed halos into larger and larger structures, the central massive galaxies are then surrounded by a number of satellites whose abundance depends on the nature of DM and on the baryonic physics driving the process of galaxy formation.

CDM models tend to predict a large number density of small-mass  galaxies ($M^*\lesssim 10^{10}$), significantly higher than is observed (\citet{Fontana06}, \citet{Marchesini09}).  In particular CDM models predict faint-end slopes of the galaxy LF that are much steeper than observed unless strong feedback from star formation activity is assumed, particularly for the dwarf galaxy population (\citet{Somerville99}, \citet{Cole00}, \citet{Menci02}). The feedback should be strong enough to inhibit all star formation in many small CDM halos as recently suggested by \citet{Sawala15}. In this case, simple abundance matching between simulated DM halos and observed galaxies would not apply. 

Alternatively, the power of the DM spectrum at small scales can be reduced if density perturbations are damped below some characteristic scale by the action of the free streaming of WDM particles with thermal relic mass of the order of $\sim 1-3$ keV, see e.g., \citet{Sommer01}. This would decrease the predicted number density of small DM halos, reducing the tension with the observed dwarf number density.  Note that 
the  suppression of the power spectrum with respect to the CDM case directly depends on the particle mass $m_X$ if the candidate is a thermal relic. For sterile neutrinos, the same suppression is obtained for a mass $m_{sterile}\approx 4.3\,m_X$ if the sterile neutrinos are produced from oscillations with active neutrinos in the \citet{Dodelson94} scenario (see \citet{Viel05}). For different production mechanisms (see \citet{Kusenko09}) the conversion factor $m_{sterile}/m_X$ ranges between 2 and 4  (see \citet{Destri13}).

Here we discuss the potential implications of our findings, and especially the  constraints on the mass of the WDM particles. In particular, we compare the faint-end slope of the LF shown in Figure 5 with those derived from the sub-halo mass function in different DM models, including a proper treatment of the present uncertainty on the  $M/L$ ratio of dwarf galaxies. The mass dependence of the cumulative mass function can be described by a universal form $N(>M)/N_{tot}\propto \big[M/M_{host}\Big]^{\alpha_{DM}}$, independent of the mass of the host structure $M_{host}$, with an exponent determined by the assumed power spectrum. CDM simulations yield $\alpha_{DM}=-1.9$ (see also \citet{Klypin11}), while smaller slopes are derived for WDM particles 
with mass $m_X\approx 1$ keV (see, e.g., Knebe et al. 2005; Maccio \& Fontanot 2010; Lovell et al. 2014). 

To derive the slope for different WDM models we need to explore the effects of different power spectra 
(corresponding to different DM particle masses) on the 
dynamical evolution of sub-halos. While full high-resolution N-body simulations provide complete theoretical tools to perform such computations, they are also extremely time consuming and are generally applied to study 
specific DM models. In addition, they are limited in resolution to sub-halo masses $M/M_{host}\gtrsim 10^{-4}$. To effectively explore a wide range of DM (thermal relic) candidate masses 
($m_X=1$, $m_X=1.5$, $m_X=2$ keV, and the CDM case $m_X\gg 1$ keV) with a resolution extending down to $M/M_{host}\approx 10^{-5.5}$, we adopt a semi-analytic model 
(see \citet{Somerville14} for a review of N-body and semi-analytic models as theoretical tools to study galaxy formation). 

This model (see \citet{Menci12}, \citet{Menci14} for details)  follows the merging histories of DM halos and sub-halos through a Monte Carlo simulation of the collapse and subsequent merging history of the peaks of the primordial density field. This lets us generate a synthyetic catalogue of model DM halos and of their past merging histories. Although the model can include a specific treatment of the baryonic physics inside the halo's evolution (e.g., gas cooling, star formation etc.), we prefer to derive constraints on the WDM particle mass assuming a general parametric power-law dependence between the galaxy luminosity and the total DM mass.
The model includes the main processes affecting the dynamical evolution of  sub-haloes within larger DM haloes, i.e., dynamical friction, merging, and stripping . 
For a recent comparison of the dynamics of sub-halos in semi-analytic models and N-body simulations see  \citet{Somerville14} and \citet{Pullen14}. 

We first tested our semi-analytic modeling of the sub-halo population against previous works in specific 
test cases where detailed results are available. 
In the left panel of Figure 6 we have compared the sub-halo mass function obtained from our model with those obtained by \citet{Pullen14} for a  CDM spectrum and for a WDM spectrum corresponding to a thermal relic particle mass $m_X=1.5$ keV. Those authors adopt a semi-analytic model calibrated and tested through the comparison with ultra high-resolution N-body simulations. The average slopes of the sub-halo mass functions we derive match the existing results  not only  for the CDM case (black curve, black points) but also for the WDM case (red dotted curve, red points) with $m_X=1.5$ keV.

As we increase the relic WDM particle masses in our model, we obtain increasing slopes for the DM sub-halo mass distribution.
In particular we obtained $\alpha_{DM}=-1.2$ for $m_X=1$ keV, $\alpha_{DM}=-1.4$ for $m_X=1.5$ keV,  $\alpha_{DM}=-1.6$ for $m_X=2$ keV, and  $\alpha_{DM}=-2$ for the CDM case.

To convert the sub-halo mass distribution $N(M)$ into a LF (i.e. a $N(L)$) we parametrize the mass-to-light ratio at low masses as $M/L\propto M^{1-\beta}$ (\citet{Moster13}), so that the slope of the LF can be written as $\alpha=(1+\alpha_{DM}-\beta)/\beta$ assuming proportionality between the galaxy luminosity and stellar mass. This turns out to be a good approximation for a limited range of (faint) magnitudes and for fairly old stellar populations. Both assumptions are satisfied here according to our SED fitting. Since at present substantial uncertainty affects the measured $M/L$ ratio for dwarf galaxies, our strategy is based on bracketing the present uncertainty by exploring values of $\beta$ in a wide range from 1 to 3. A value of $\beta> 1$ is derived from abundance matching techniques (\citet{Moster13}; see also \citet{Behroozi13}) assuming a standard cuspy (e.g. Navarro, Frenk \& White) inner density CDM profiles for the observed satellites in the Local Group. In this case only a fraction of CDM halos can host galaxies. More complex and shallower density profiles assuming a dynamical interaction between baryons and DM halos are consistent with a shallower slope $\beta\sim 1$ (see \citet{Brook15} and \citet{Sawala15} for both cases).

Assuming that all halos host a galaxy,  we can then derive the slope $\alpha$ of the LF obtained in different CDM/WDM scenarios as a  function of the assumed value for $\beta$, and compare it with our measured value $\alpha=-1.4\pm 0.05$ derived from our dwarf sample. The output of this exercise is shown in the right panel of Figure 6. We confirm that the CDM model needs a relatively large value of $2<\beta<2.5$ to be compatible with the data. In the case of WDM models, we are able to set a lower limit of $m_X\geq 1.5$ keV to be consistent with the measured slope $\alpha$, adopting lower values of $1<\beta<2$. On the contrary, WDM models with $m_X< 1.5$ keV are not able to provide the observed faint-end LF slope if $\beta=1$ represents a reasonable lower limit for the $M/L$ relation. Of course, our low limit on the WDM particle mass could even increase if we allow for a complete inhibition of star formation in a fraction of the low-mass sub-halo population due to reionization and background UV radiation,  as suggested for instance by \citet{Sawala15}.

This limit compares well with other existing limits. Lower limits $m_X\geq 1$ keV have been derived from galaxy counts in the Hubble Ultra Deep Field (\citet{Schultz14}), while limits $m_X\gtrsim 1.5$ keV  have been derived from the abundance of ultra-faint Milky Way satellites measured in the Sloan Digital Sky Survey (see \citet{Horiuchi14} and references therein). Much tighter constraints, $m_X\gtrsim 3.3$ keV  (e.g. \citet{Viel13}), have been derived by comparing the observed Lyman-$\alpha$ forest of high- resolution ($z > 4$) quasar spectra with hydrodynamical N-body simulations. Various uncertainties may still affect these constraints (see discussions in \citet{Abazajian11}, \citet{Schultz14}). The comparison of subhalos to Milky Way dwarfs assumes a factor $\approx  4$ correction for the number of dwarfs being missed by current surveys, and lower correction factors would appreciably weaken the constraints. Lyman-$\alpha$ absorption is also a challenging tool and requires disentangling the effects of pressure support and thermal broadening from those caused by DM spectrum, as well as assumptions on the thermal history of the intergalactic medium and of the ionizing background. 

\section{Summary}
We have used deep LBC exposures obtained  in five bands (UBriz) over a  $576$ arcmin$^2$ pointing located within the Virgo field to find and study faint LSB dwarfs associated with the Virgo cluster at an unprecedented depth of $M_r\simeq -9$. Despite the relatively small field of view, we detect in the $r$ band as many as 11 LSB dwarfs in the magnitude interval $-13\lesssim M_r\lesssim -9$, which we identify as Virgo members on the basis of their position in the $\mu_0 - r$ plane. 

Due to the significant depth of our images, their morphology appears to be contaminated by smaller, or compact sources, most likely in the background, which would affect the estimate of their total flux and morphological parameters. To remove them from our analysis, we have used Galfit to separately model the LSB dwarfs and the other sources, in order to derive contamination-free morphological parameters and magnitudes. This procedure has been applied to all images in five bands. 
We find that our sources  have scale radii $250\lesssim r_s\lesssim 850$ pc and an average red color $U-r\sim 2.1$. 
Based on this evidence, we claim that these sources represent the faintest extremes of the population of low luminosity LSB dwarfs that has been recently detected in wider surveys of the Virgo cluster (e.g. \citet{Trentham02}, \cite{Sabatini03}, \citet{Lieder12} and \citet{Davies15}).  

The average ratio between the age of the last episode of star formation and the exponentially declining star formation timescale $\langle age/\tau\rangle=3.6$ indicates a passively evolving phase with a specific SFR of $\sim 10^{-11}$ yr$^{-1}$. They are older ($\sim 7$ Gyr) than the typical cluster formation age and appear regular in morphology.

We have converted our detections into  average projected number densities,  normalized at 200 kpc from the Virgo center, computed in two magnitude bins centered at $\langle M_r\rangle =-12.5$ and $\langle M_r\rangle =-10.5$. We find a number density of 320 and 560 sources Mpc$^{-2}$ mag$^{-1}$, respectively. These results allow us to extend the estimate of the Virgo LF down to very faint limits.  Merging our data with the brighter points mainly based on the VCC survey, we find an average faint-end slope of about $\alpha \simeq -1.4$ down to $M_r\sim -9.5$.

It is well known  that the relatively shallow slope of the LF sets challenging constraints on Dark Matter scenarios. Strong baryonic feedback processes must be invoked to reconcile the intriniscally steep CDM slopes with the observed  flatter LF. 
Instead, we explore  the alternative options, i.e., those provided by WDM scenarios with particles of $\simeq $keV mass scale.

Using our semi-analytic model \citep{Menci12} we show that the slope $\simeq -1.4$ we derive for the LF can be translated into an interesting low limit on the WDM particle mass $>1.5$ keV, regardless of the exact relation present at low masses between the $M/L$ ratio and the halo mass $M$ (i.e. regardless of the details of the baryonic feedback processes). This limit compares well with previous limits derived from galaxy counts in the Hubble Ultra Deep Field (\citet{Schultz14}), from the abundance of ultra-faint Milky Way satellites in the Sloan Digital Sky Survey (see \citet{Horiuchi14}) or from the observed Lyman-$\alpha$ forest  (e.g., \citet{Viel13}).

It is intriguing to see that this small but non-negligible sample has been detected in a single pointing taken with a $\simeq 24' \times 24'$ imager at an 8m telescope (LBC at LBT in this case). While it is still limited by the small area surveyed so far and by the lack of spectroscopic follow up, the extension of the same analysis to a larger Virgo region using deep (2-4h of exposure time per band) multicolor images at 8m class telescopes is clearly needed to reduce the current uncertainties.

\acknowledgments
We thank the referee for useful comments which have significantly improved our paper. We also thank M. Dickinson for critical reading. Observations have been carried out using the Large Binocular Telescope at Mt. Graham, AZ. The LBT is an international collaboration among institutions in the United States, Italy, and Germany. LBT Corporation partners are The University of Arizona on behalf of the Arizona university system; Istituto Nazionale di Astrofisica, Italy; LBT Beteiligungsgesellschaft, Germany, representing the Max-Planck Society, the Astrophysical Institute Potsdam, and Heidelberg University; The Ohio State University; and The Research Corporation, on behalf of The University of Notre Dame, University of Minnesota, and University of Virginia.


{}

\clearpage

\begin{table}[tb]\footnotesize
\caption{LSB catalog in the Virgo-xmmuj1230 field}
\begin{center}
\begin{tabular}{c r r r r r r r r r r r}
\hline\hline
ID & A\,\,\,\,\, & B\,\,\,\,\, & C\,\,\,\,\,  & D\,\,\,\,\, & E\,\,\,\,\, & F\,\,\,\,\, & G\,\,\,\,\, & H\,\,\,\,\, & J\,\,\,\,\,  & K\,\,\,\,\, & L\,\,\,\,\,  \\
\hline
RA (187) & 0.3919 & 0.6635 & 0.5879 & 0.5483 & 0.3814 & 0.3863 & 0.6600 & 0.6224 & 0.4138 & 0.7070 & 0.7244 \\
DEC (+13) & 0.7704 & 0.7391 & 0.7058 & 0.6914 & 0.6619 & 0.6222 & 0.6196 & 0.5565 & 0.5075 & 0.5053 & 0.4939 \\
$r$ & 21.8 & 20.4 & 21.2 & 18.3 & 18.8 & 21.9 & 18.1 & 21.2 & 21.2 & 19.1 & 21.9 \\
$\mu_0$ & 24.9 & 24.5 & 25.8 & 24.0 & 23.5 & 24.5 & 24.3 & 25.2 & 25.5 & 25.8 & 26.1 \\
$n^a$ & 0.9 & 0.7 & 0.7 & 0.7 & 0.7 & 0.9 & 0.6 & 0.5 & 0.7 & 0.4 & 0.3 \\
M$_R^b$ & -9.3 & -10.7 & -9.9 & -12.8 & -12.4 & -9.3 & -13.1 & -9.9 & -9.9 & -12.0 & -9.3 \\
r$_s$ (arcsec)$^c$ & 3.1 & 4.5 & 6.1 & 8.0 & 5.7 & 2.7 & 10.4 & 3.0 & 4.2 & 10.7 & 3.0 \\
r$_s$ (pc)$^c$  & 248 & 360 & 488 & 640 & 456 & 216 & 832 & 240 & 336 & 856 & 240 \\
axis ratio & 0.7 & 0.7 & 0.7 & 0.9 & 0.7 & 0.6 & 0.7 & 0.9 & 0.8 & 0.9 & 0.9 \\
$r-i$ & 0.1 & 0.1 & 0.2 & 0.2 & 0.1 & 0.1 & 0.1 & 0.3 & 0.2 & 0.0 & 0.0 \\
$i-z$ & 0.2 & 0.2 & 0.0 & 0.2 & 0.2 & 0.0 & 0.2 & 0.2 & 0.0 & -0.2 & 0.1 \\
$B-r$ & 0.8 & 0.9 & 0.7 & 1.0 & 0.9 & 0.9 & 0.9 & 0.8 & 0.9 & 0.8 & 0.7 \\
$U-B$ & 0.8 & 1.0 & 1.4 & 1.2 & 1.0 & 1.2 & 1.1 & 1.2 & 1.5 & 2.2 & 0.8 \\
SFR$^d$  & 2 & 5 & 2 & 17 & 22 & 1 & 30 & 2 & 0.1 & 37 & 4 \\
M$^{*e}$  & 0.5 & 2 & 1 & 20 & 10 & 0.5 & 22 & 1 & 1 & 4 & 0.3 \\
ages$^f$ & 7 & 7 & 7 & 9 & 7 & 7 & 8 & 8 & 8 & 6 & 6 \\
$\tau^g$ (Gyr) & 3 & 2 & 2 & 2 & 2 & 2 & 2 & 2 & 1 & 3 & 5 \\
\hline
\hline
\end{tabular}
\end{center}
$^a$ S{\'e}rsic index\\
$^b$ Absolute magnitudes computed adopting an average distance modulus for Virgo $\Delta M=31.1$ (\citet{Mei07}).\\
$^c$ Scale radius from the Sersic profile fitting. An angular scale of 80 pc arcsec$^{-1}$ has been adopted.\\
$^d$ Star formation rate in units of $10^{-5}$ M$_{\odot}$ yr$^{-1}$. Resulting best fit Bruzual \& Charlot models have no dust and 0.2 solar metallicity. SFR are uncertain by a factor 1.2 on average.\\
$^e$ Stellar mass in units of $10^6$M$_{\odot}$. Average uncertainties are of the order of 40\%.\\
$^f$ Ages in units of Gyr. The 68\% average probability distribution ranges from 6 to 12 Gyr.\\
$^g$ Exponential timescale of declining star formation. The 68\% average probability distribution ranges from 2 to 8 Gyr.
\end{table}

\clearpage

\begin{figure*}
\centering
\scalebox{0.6}[0.6]{\rotatebox{0}{\hspace{0cm}\includegraphics{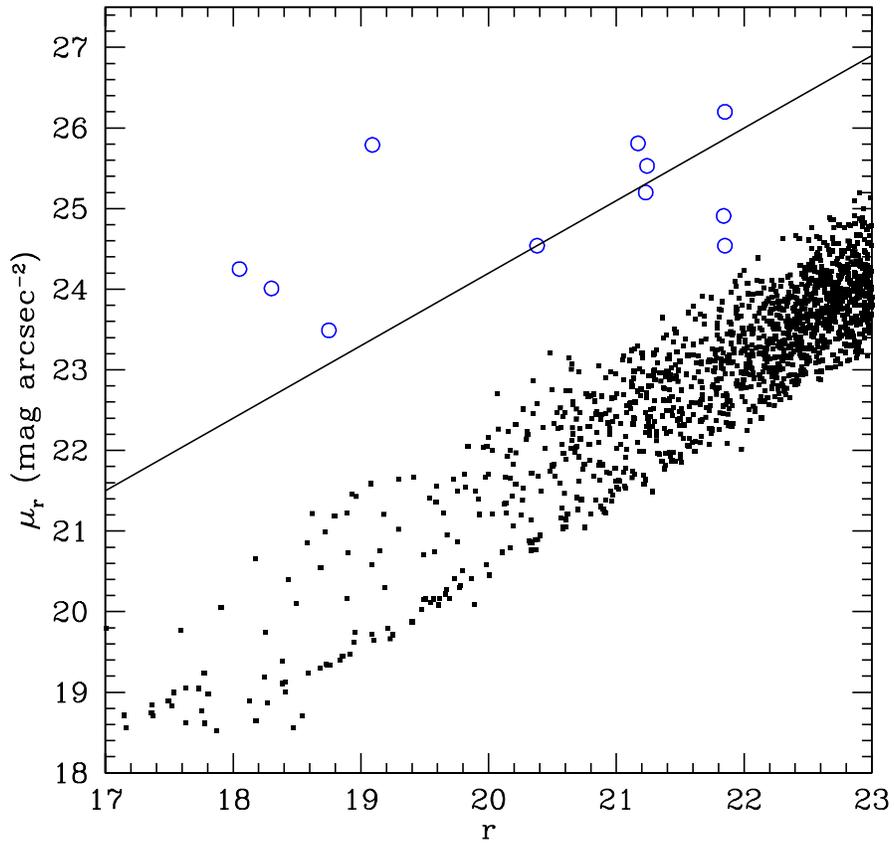}}}
\caption{Surface brightness - magnitude relation for the galaxies in the field. Faint sources located in the extended halos of bright galaxies and blended galaxies have been removed from the plot since they have biased photometry. LSB dwarf candidates are shown as blue circles.
The continuous line represents an extrapolation to fainter magnitudes of the LSB selection threshold adopted by \citet{Rines08}}
\end{figure*}

\clearpage
\begin{figure*}
\centering
\scalebox{0.6}[0.6]{\rotatebox{0}{\hspace{0cm}\includegraphics{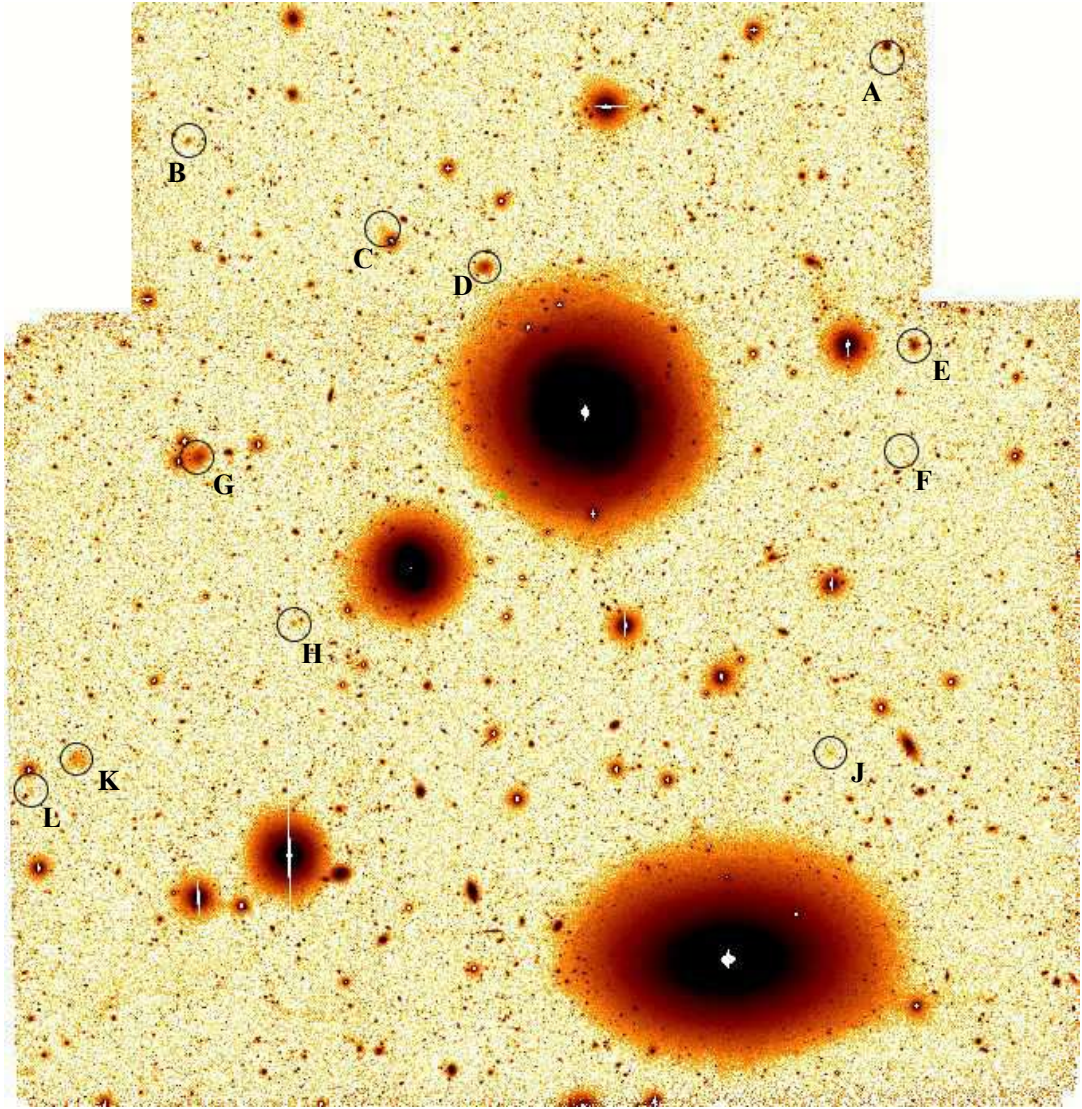}}}
\caption{Position of the selected LSB dwarfs in the Virgo-xmmuj1230 LBC field}

\end{figure*}

\clearpage
\begin{figure*}
\centering
\scalebox{0.6}[0.6]{\rotatebox{0}{\hspace{0cm}\includegraphics{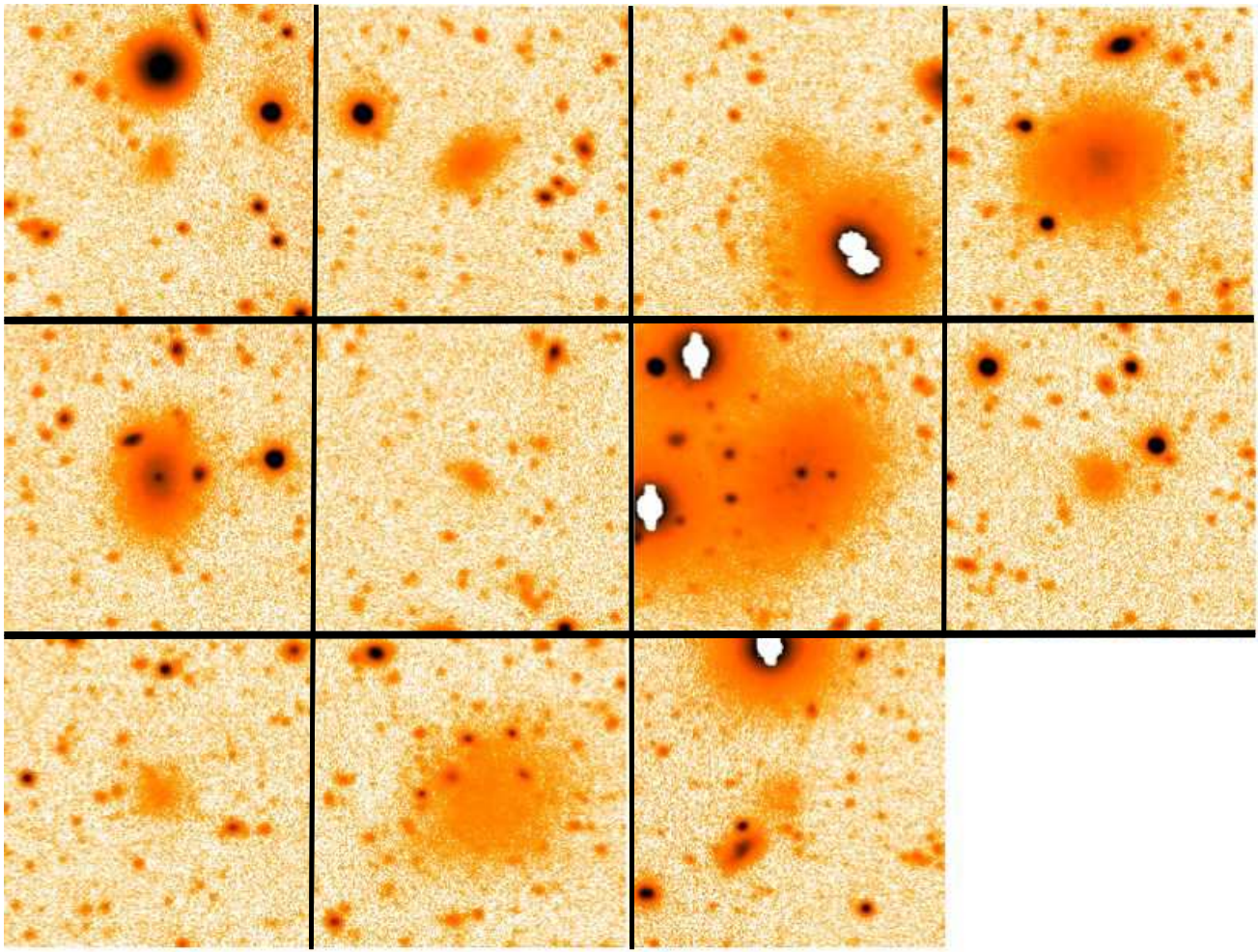}}}
\caption{Selected LSB dwarfs in the Virgo-xmmuj1230 field; the box size of each image is $\simeq 57$ arcsec. The sequence from the top-left to the bottom follows the list in Table 1.}
\end{figure*}

\clearpage
\begin{figure*}
\centering
\scalebox{0.6}[0.6]{\rotatebox{0}{\hspace{0cm}\includegraphics{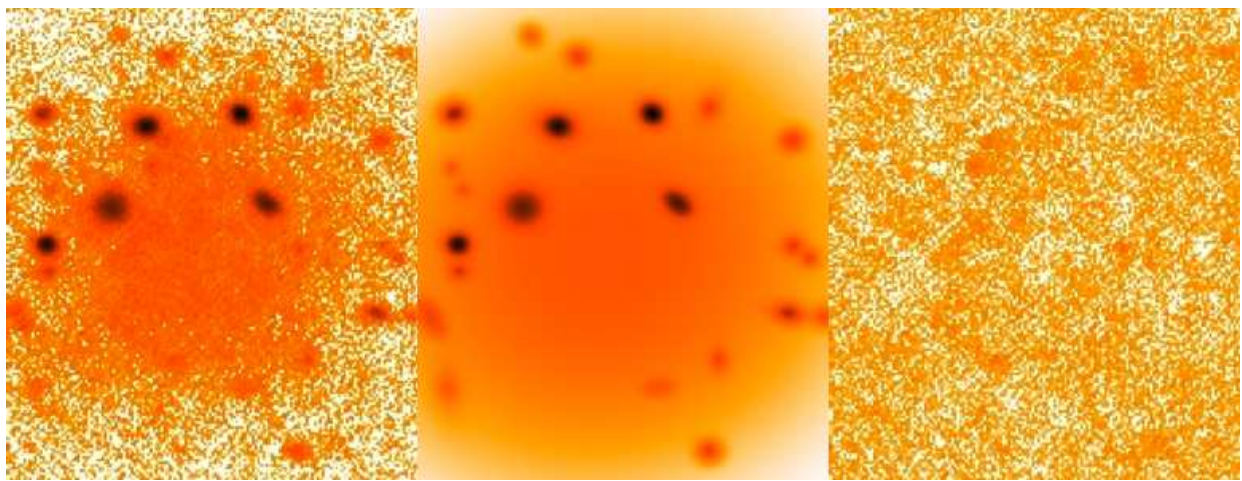}}}
\caption{The "K" dwarf galaxy. From left to right: observed profile; model profile from Galfit (the brightest background galaxies have been fitted separately); residuals.}
\end{figure*}

\clearpage
\begin{figure*}
\centering
\scalebox{0.6}[0.6]{\rotatebox{0}{\hspace{0cm}\includegraphics{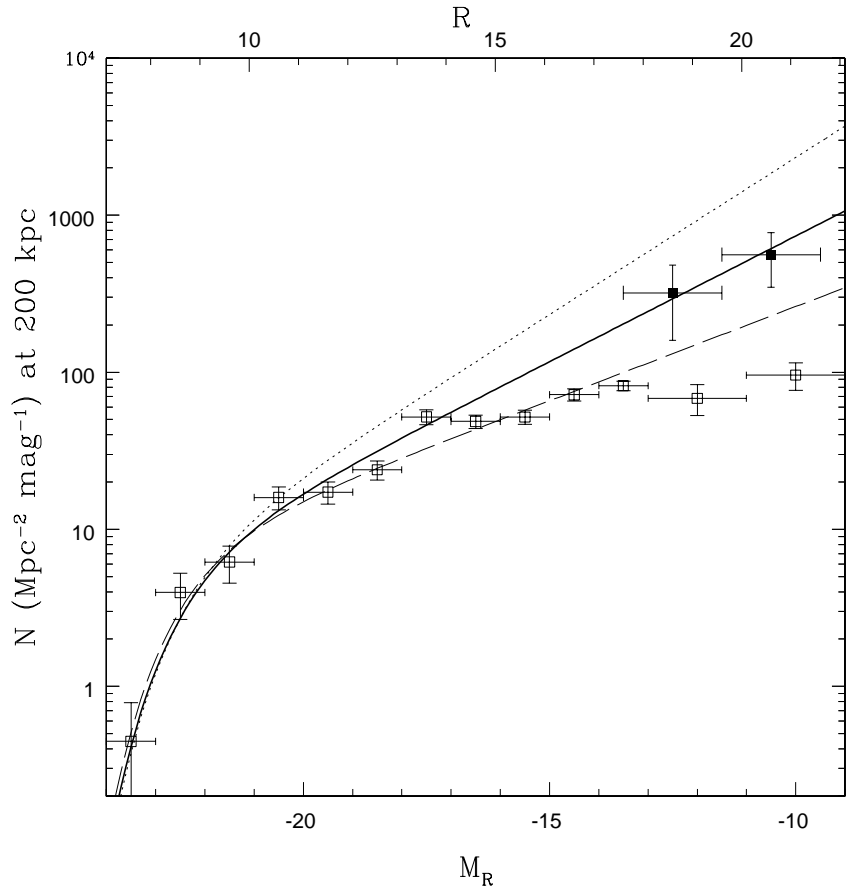}}}
\caption{Virgo projected LF normalized at 200 kpc. Filled squares indicate data from the present sample after conversion from AB to the Vega magnitude system by $ R(Vega)=r(AB)-0.2$. Empty squares are from  \citet{Trentham02} in the Vega system. The continuous curve represents a Schechter shape with slope $\alpha \sim -1.4$ and $M^*\sim -22.5$. Two faint slopes $\alpha \sim -1.3,-1.5$ are also shown for comparison (dashed and dotted lines, respectively).}
\end{figure*}

\clearpage

\clearpage
\begin{figure*}
\centering
\scalebox{0.7}[0.7]{\rotatebox{-90}{\hspace{0cm}\includegraphics{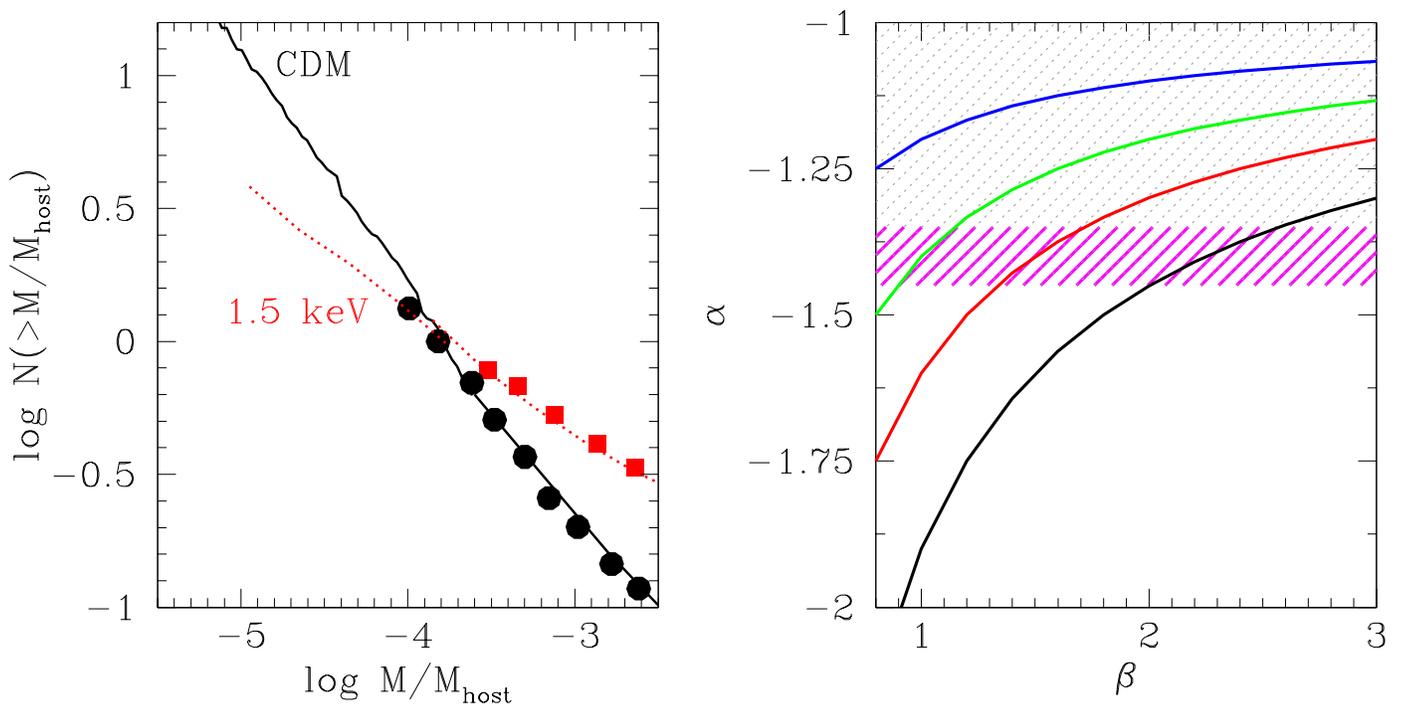}}}
\caption{DM model predictions. Left: Cumulative mass function vs satellite over main halo mass ratio predicted by our Monte Carlo model (black and red curves) compared with results derived  by \citet{Pullen14}. Black curve and points are for the CDM spectrum and red curve and points for WDM spectrum with $m_X=1.5$ keV. Right: luminosity (i.e. stellar mass)  function slope $\alpha$ vs  $\beta$, the slope of the $M/L\propto M^{1-\beta}$  power-law relation. The observed value $\alpha=-1.4\pm 0.05$ is shown as a dashed region. Different curves derived from our semi-analytic model are shown for different WDM particle masses and CDM (1,1.5,2 keV and CDM starting from top to bottom). For $\beta>1$ then $m_X>1.5$ keV if $\alpha\simeq -1.4$.  }
\end{figure*}

\end{document}